\documentclass[aps,10pt,twocolumn,superscriptaddress,preprintnumbers,amsmath,amssymb,prl]{revtex4-1}

\usepackage{times}
\usepackage[pdftex]{graphicx}
\usepackage[bookmarksopen=true,bookmarksopenlevel=1]{hyperref}
\usepackage{epstopdf}
\usepackage{amsmath}
\usepackage{amsfonts}
\usepackage{amssymb}
\usepackage{graphicx}
\usepackage{sidecap}
\sidecaptionvpos{figure}{c}
\usepackage{color}
\usepackage{dcolumn}
\usepackage{bm}
\usepackage[compact]{titlesec}
\usepackage[normalem]{ulem}
\usepackage{bookmark}
\usepackage{natbib}
\usepackage{indentfirst}
\usepackage{appendix}

\definecolor{mypink}{RGB}{219, 48, 122}
\definecolor{mygreen}{RGB}{51, 153, 102}
\definecolor{brown}{RGB}{165, 42, 42}

\definecolor{mypink}{RGB}{219, 48, 122}
\definecolor{mygreen}{RGB}{51, 153, 102}
\definecolor{brown}{RGB}{165, 42, 42}

\newlength{\beforesection}
\newlength{\aftersection}
\newlength{\beforesubsection}
\newlength{\aftersubsection}
\titlespacing*{\section}{0pt}{\beforesection}{\aftersection}
\titlespacing*{\subsection}{0pt}{\beforesubsection}{\aftersubsection}

\def\root33{$\sqrt{3}\times\sqrt{3}$ {\it R}30$^\circ$}
\def\RT3{$\sqrt{3}$}

\def\MnSbSbTe4{MnSb$_2$Te$_4$}
\def\MnBiBiTe4{MnBi$_2$Te$_4$}
\def\MnBiBiSe4{MnBi$_2$Se$_4$}

   \makeatletter
\def\@fnsymbol#1{\ensuremath{\ifcase#1\or \dagger\or  \ast\or
   \mathsection\or \mathparagraph\or \|**\or \or \ast\ast
   \or \ast\ast \else\@ctrerr\fi}}
    \makeatother
    
\usepackage{mathtools}

\DeclarePairedDelimiterX\braket[2]{\langle}{\rangle}{#1 \delimsize\vert #2}

\def\root33{$\sqrt{3}\times\sqrt{3}$ {\it R}30$^\circ$}
\def\RT3{$\sqrt{3}$}

\def\MnSbSbTe4{MnSb$_2$Te$_4$}
\def\MnBiBiTe4{MnBi$_2$Te$_4$}
\def\MnBiBiSe4{MnBi$_2$Se$_4$}

   \makeatletter
\def\@fnsymbol#1{\ensuremath{\ifcase#1\or \dagger\or  \ast\or
   \mathsection\or \mathparagraph\or \|**\or \or \ast\ast
   \or \ast\ast \else\@ctrerr\fi}}
    \makeatother

\begin{document}

\title{Hidden spin-orbital texture at the \texorpdfstring{$\overline{\Gamma}$}{TEXT}-located valence band maximum of a transition metal dichalcogenide semiconductor}

\author{Oliver J. Clark}
\email[Corresponding author. E-mail address: ] {oliver.clark@helmholtz-berlin.de}
\affiliation{Helmholtz-Zentrum Berlin f\"ur Materialien und Energie, Elektronenspeicherring BESSY II, Albert-Einstein-Str. 15, 12489 Berlin, Germany}

\author{Oliver Dowinton}
\affiliation{Department of Physics and Astronomy, University of Manchester,
Oxford Road, Manchester M13 9PY, United Kingdom}

\author{Mohammad Saeed Bahramy}
\affiliation{Department of Physics and Astronomy, University of Manchester,
Oxford Road, Manchester M13 9PY, United Kingdom}

\author{Jaime S\'anchez-Barriga}
\affiliation{Helmholtz-Zentrum Berlin f\"ur Materialien und Energie, Elektronenspeicherring BESSY II, Albert-Einstein-Str. 15, 12489 Berlin, Germany}
\affiliation{IMDEA Nanoscience, C/ Faraday 9, Campus de Cantoblanco, 28049, Madrid, Spain}

\begin{abstract} 

Finding stimuli capable of driving an imbalance of spin-polarised electrons within a solid is the central challenge in the development of spintronic devices. However, without the aid of magnetism, routes towards this goal are highly constrained with only a few suitable pairings of compounds and driving mechanisms found to date. Here, through spin- and angle-resolved photoemission {along with density functional theory}, we establish how the $p$-derived bulk valence bands of semiconducting 1T-HfSe$_2$ possess a local, ground-state spin texture spatially confined within each Se-sublayer due to strong sublayer-localised electric dipoles orientated along the $c$-axis. 
This hidden spin-polarisation manifests in a `coupled spin-orbital texture' with in-equivalent contributions from the constituent $p$-orbitals. While the overall spin-orbital texture for each Se sublayer is in strict adherence to time-reversal symmetry (TRS), {spin-orbital mixing terms with net polarisations at time-reversal invariant momenta are locally maintained. These apparent TRS-breaking contributions dominate,} and can be selectively tuned between with a choice of linear light polarisation, facilitating the observation of pronounced spin-polarisations at the Brillouin zone centre for all $k_z$. We discuss the implications for the generation of spin-polarised populations from 1T-structured transition metal dichalcogenides using a fixed energy, linearly polarised light source.
\end{abstract}

\maketitle

\section{INTRODUCTION}

Spintronic devices aim to exploit the spin quantum number of an electron rather than the charge~\cite{hirohata_review_2020}. Their operational principles are underpinned by a reversible external stimulus to which electrons of opposing spin species respond oppositely~\cite{sierra_van_2021}. In non-magnetic systems  the task of finding such stimuli is not straightforward due to the
presence of time-reversal symmetry ($\text{TRS}: E(k,\uparrow)$ = $E(-k, \downarrow)$ where $E$, $\pm k$, and $\uparrow \downarrow$ denote the electron energy, momentum and spin respectively), enforcing a net-zero spin polarisation for all electronic bands across a material. One therefore must enforce an imbalance in spin species across the material by applying a magnetic field~\cite{datta_electronic_1990,yamada_electrically_2011}, or by selectively coupling to electrons at $\pm k$ unevenly, in addition to the selective coupling to a single spin species~\cite{kato_current_2004, zeng_valley_2012, mak_control_2012}. The latter, all-electronic pathway permits easier integration with traditional components, and is thus more desirable~\cite{kato_current_2004, aswschalom_spintronics_2009}, but only a few known compounds host the necessary spin textures within their electronic structure along with an in-built mechanism to couple from opposite $k$ vectors reliably. 

Of these compounds, several belong to the transition metal dichalocgenide (TMD) material family. TMDs consist of van der Waals separated layers of X-M-X formula units (X$\in \{$S, Se, Te$\}$), with the transition metal atom (M)  positioned at the centre of each MX$_6$ octahedron~\cite{chhowalla_chemistry_2013}. 
The TMDs are renowned for their array of often-overlapping superconducting~\cite{allan_high_1998, ryu_superconductivity_2015, dvir_spectroscopy_2018, clark_fermiology_2018}, charge density wave~\cite{wilson_charge_1975, arguello_visualizing_2014, rossnagel_fermi_2005, borisenko_two_2009, watson_orbital_2019} and topological~\cite{li_topological_2017, bahramy_ubiquitous_2018, nicholson_uniaxial_2021, ghosh_observation_2019, mukherjee_fermi_2020, clark_fermiology_2018} phases, but the modern interest in this series stemmed from the similarity between their two-dimensional honeycomb lattice structures to that of graphene. Indeed, the electronic structure of monolayer 1H-MoS$_2$ and 1H-WSe$_2$ can be derived from that of graphene by breaking A-B sublattice symmetry and increasing the atomic spin-orbit coupling strength~\cite{mak_atomically_2010}. The resulting pair of $d_{xy}$ and $d_{x^2-y^2}$-derived spin-split valence bands at the K  and K' points of the bulk Brillouin zone couple differently to circularly polarised light, enabling the generation of spin-polarised currents in the conduction band and prompting the design of so-called valleytronic devices exploiting the spin-valley coupling~\cite{xaio_coupled_2012, zeng_valley_2012, mak_control_2012, yuan_zeeman_2013}.

In their bulk forms, MoS$_2$ and WSe$_2$ adopt an inversion symmetric ($\text{IS:}  E(k,\uparrow)$ = $E(-k, \uparrow)$) 2H structure.  
While the combination of IS and TRS enforce that the bulk band structure is entirely spin-degenerate, a so-called `hidden spin polarisation' is still directly observable at the K and K' points with a surface sensitive probe of the electronic structure due to a strong localisation of the  $d_{xy}$ and $d_{x^2-y^2}$ wavefunctions to within a single 1H-sublayer~\cite{riley_direct_2014, bawden_spin_2015, razzoli_selective_2017}, with similar phenomena observable in other centrosymmetric compounds~\cite{zhang_hidden_2014, guan_electrically_2020, gatti_hidden_2021}. However, the transition to the 2H-TMD structure shifts the valence band maximum from the K point to the time-reversal invariant momentum (TRIM), $\Gamma$, complicating the generation of clean spin-currents in the bulk compounds~\cite{splendiani_emerging_2010, ellis_indirect_2011,zhang_direct_2014}.

Here, we search for an analogous hidden spin mechanism in 1T-TMDs. Through spin- and angle-resolved photoemission (spin-ARPES), we show that time- and inversion-symmetric 1T-HfSe$_2$ exhibits a hidden spin-polarisation in the $\overline{\Gamma}$-centred Se $p$-derived valence bands originating from the asymmetry of a single 1T unit cell along the $c$-axis. 
Remarkably, though, we show how this spin-polarisation persists through the $\overline \Gamma$ point of the surface Brillouin zone and retains the same sign for $\pm k$ vectors. {Our density functional theory (DFT) calculations show how} this effective net spin-polarisation is of entirely non-magnetic origin, instead deriving from  {local spin-orbital mixing within the $p$-orbital manifold, aided by} a selective coupling of linearly polarised light to subsets of $p$-orbitals, and therefore to a partial ground state spin-texture. 

\section{RESULTS AND DISCUSSION}

\noindent{\textbf{Hidden spin mechanism in 1T-structured TMDs}}

\begin{figure}[h]
	\centering
	\includegraphics[width=0.8\columnwidth]{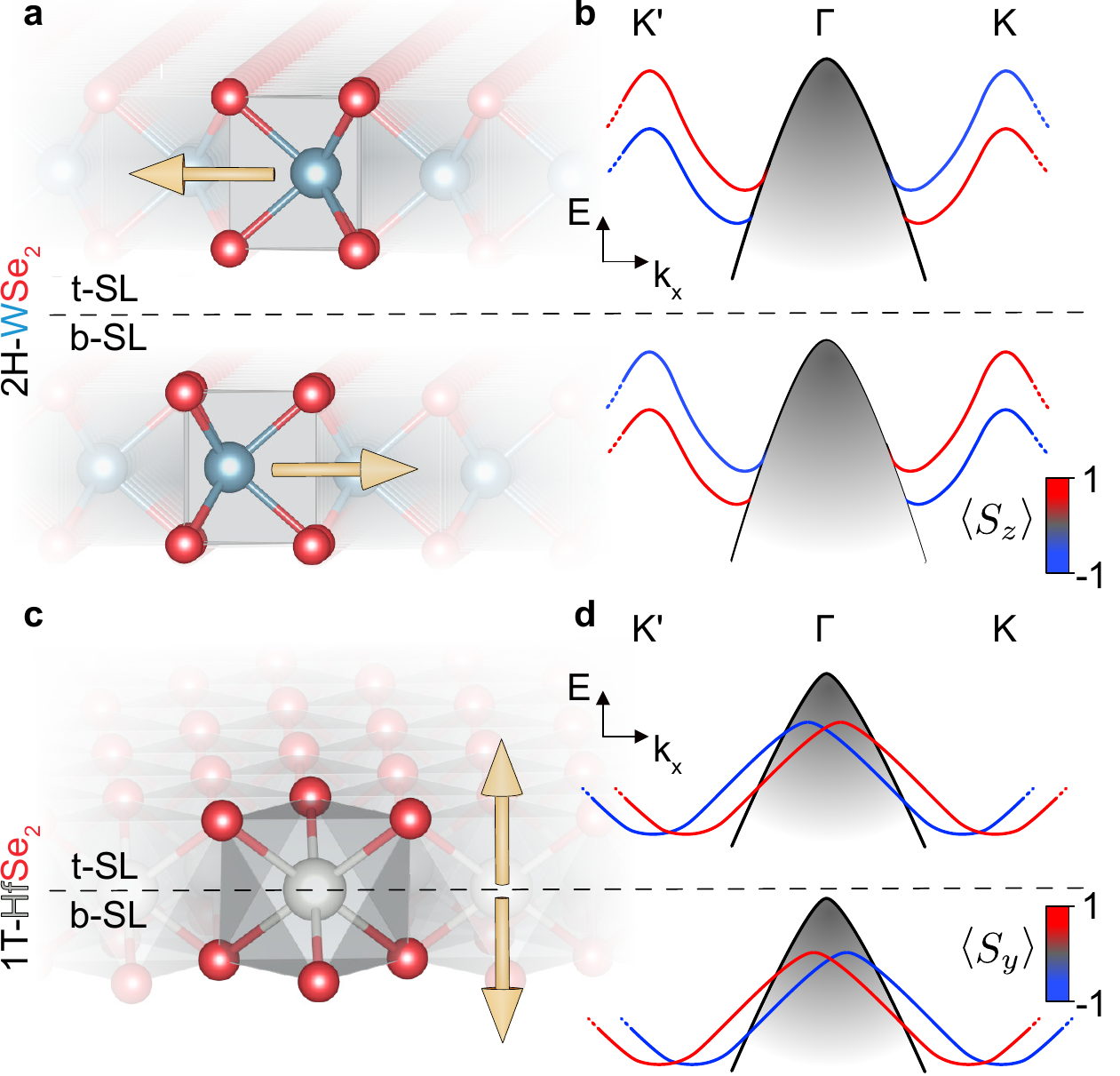}
	\caption{\label{Fig1} \textbf{Hidden spin in TMDs} (a) Real-space structure of 2H-WSe$_2$. A single unit cell is highlighted. Dashed line indicates the division between top (t) and bottom (b) sublayers (SL). Arrows indicate the direction of the effective electric field originating from W-Se dipoles. (b) Schematic of the sub-layer resolved valence band structure of bulk 2H-WSe$_2$. Colours indicate the direction of the out-of-plane Rashba-type spin polarisation; (c-d) Equivalent schematics for 1T-HfSe$_2$. The effective out-of-plane dipole originating from Hf-Se bonds is localised to a half unit cell, giving rise to an in-plane Rashba splitting of the $p_{x,y}$ bands around the $\Gamma$ point. 
	}
\end{figure}

Figure~\ref{Fig1} compares the hidden spin mechanism in 2H and 1T structured bulk TMDs using 2H-WSe$_2$ and 1T-HfSe$_2$ as examples. In the 2H structure, the triangularly coordinated X atoms either-side of the MX$_6$ octahedron are orientated identically within a single sub-layer. This causes a charge imbalance orientated parallel to the layers, breaking inversion symmetry on a local scale. 
Fig.~\ref{Fig1}(b) schematizes the resulting layer-resolved electronic structure~\cite{riley_direct_2014}. The Rashba effect produces a spin-polarisation in the directions perpendicular to any applied field~\cite{bychkov_oscillatory_1984}, here producing an out-of-plane Zeeman-like splitting in the W $d_{xy}$ and $d_{x^2-y^2}$-derived bands at the K points, reversing sign at the K' points in accordance with TRS. The spin-polarisation is exactly opposite in the second sublayer, and therefore non-zero spin polarisations are uncovered only for surface sensitive probes of electronic structure~\cite{riley_direct_2014, bawden_spin_2015}.

For the 1T-structure, the chalcogen sublayers are 180$^{\circ}$ rotated from one-another either side of the transition metal plane. Charge is thus equally distributed across a single X-M-X layer, and the 1T structure is inversion symmetric down to a single monolayer. A net dipole now exists out-of-plane, however, switching sign about the M plane. While the spatial scales for this local inversion symmetry are very small, previous studies have predicted that hidden spin physics can arise in 1T-TMDs~\cite{cheng_hidden_2018}, with a local in-plane Rashba polarisation verified in the Se $p_{x,y}$-derived bands of monolayer 1T-PtSe$_2$ by spin-ARPES~\cite{yao_direct_2017}, {in-line with that schematised in Fig.~\ref{Fig1}(d) for the top and bottom sub-layers}. For a bulk system, the presence of hidden spin polarisation is yet to be confirmed. For the present case, Hf is amongst the least electronegative of all transition metals, producing a particularly strong electric dipole across a Hf-Se bond. The degree of spin-splitting from a hidden spin mechanism should be particularly pronounced in this compound.

\

\noindent{\textbf{Overview of the electronic structure of 1T-HfSe$_2$}}

\begin{figure}[ht!]
	\centering
	\includegraphics[width=\columnwidth]{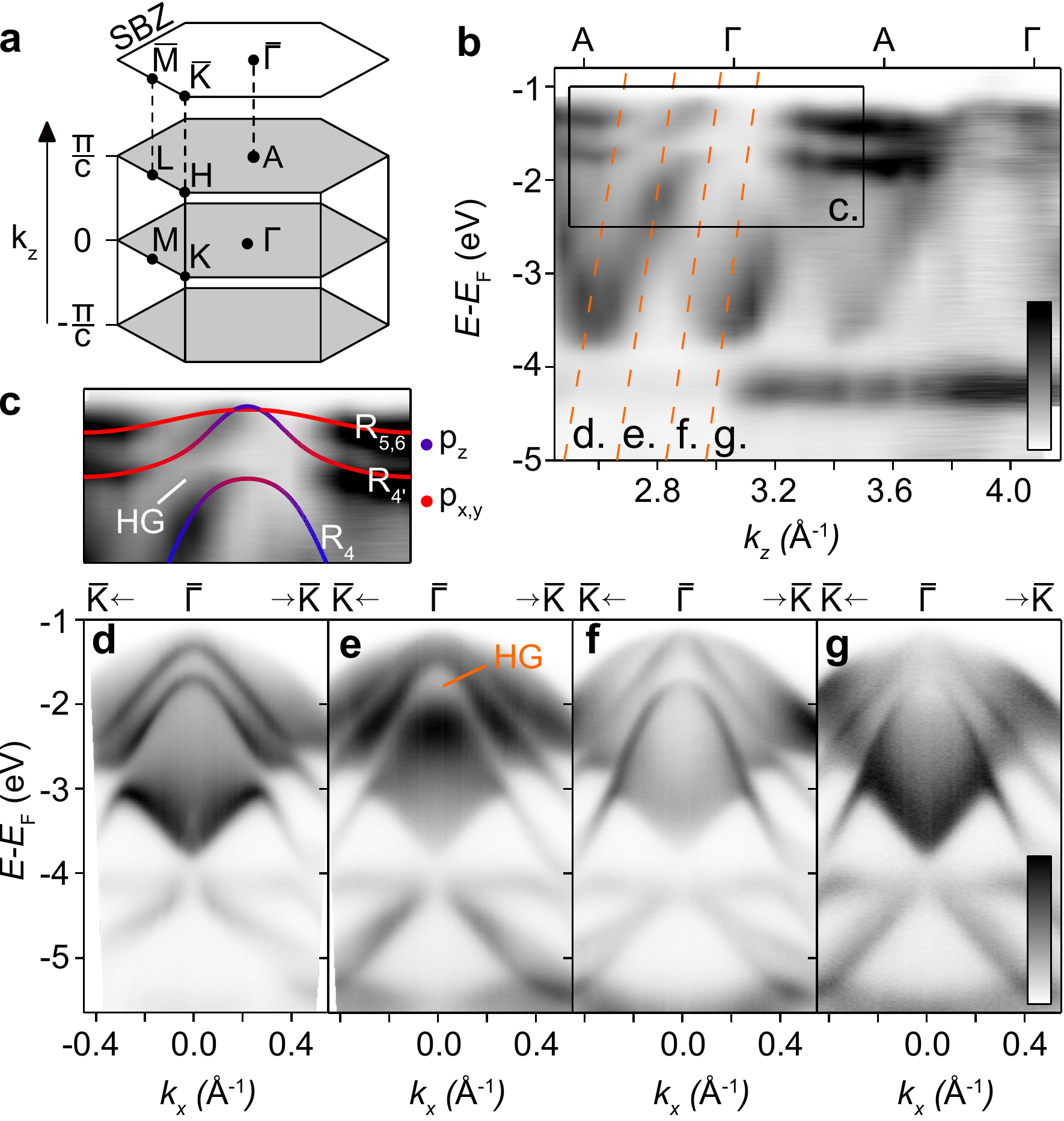}
	\caption{\label{Fig2} \textbf{Valence electronic structure of HfSe$_2$.} (a) Bulk and surface Brillouin zone (SBZ) schematics with high symmetry points labelled. The surface plane lies perpendicular to the $k_z$ direction. (b) $k_z$ dispersion formed from a photon energy dependent dataset ranging from $h\nu=$17-64~eV (see Methods). The rectangle indicates the energy and $k_z$ range for the zoom-in in (c). Orange dashed lines indicate the photon energies corresponding to the dispersions in (d-g). (c) Zoom-in over the range indicated in (b). {Solid lines are orbital-projected DFT calculations, re-scaled in energy by 25\%, and offset in ($k_z$,$E_B$) by (0.07 \AA$^{-1}$, -1.15~eV) in order to optimally match the experimental data in the low h$\nu$ region. Colour is indicated to the right of the image. The irreducible representation for each band is labelled, along with the hybridisation gap (HG).} (d-f)  Dispersions along the $\overline{\mbox{K}}$-$\overline{\Gamma}$-$\overline{\mbox{K}}$ direction for photon energies of (d) 22~eV (A-plane) (e) 25.5~eV (f) 29~eV and (g) 32~eV ($\Gamma$-plane).
	25.5~eV probes the $k_z$ where the hybridisation gap is formed. Note the lack of a visible surface state occupying the gap. }
\end{figure}

 Figure~\ref{Fig2} overviews the electronic structure of 1T-HfSe$_2$ as seen by ARPES, which probes only the occupied part of the band structure below the Fermi level, $E_{\text{F}}$. The bulk and surface Brillouin zones (BZ) are displayed in Fig.~\ref{Fig2}(a) with high symmetry points indicated.
 {The degree of atomic orbital overlap in 1T-structured TMDs is, in general, low, with the bonding (B) and anti-bonding (AB) chalcogen derived states energetically separated from the $e_g$ and $t_{2g}$ manifolds. For the group IV TMDs to which HfSe$_2$ belongs, the transition metal is in the $d^0$ configuration, with the Fermi level falling between the unoccupied $t_{2g}$ manifold, and the AB-Se $p$-derived states~\cite{chhowalla_chemistry_2013, eknapakul_direct_2018}}. HfSe$_2$ is thus a indirect gap semiconductor ($E_G \approx$ 1.1~eV~\cite{gaiser_band_2004}), with the Se-derived valence band maximum (VBM) and Hf-derived conduction band minimum (CBM) located at the $\Gamma$ and M points of the bulk BZ,  respectively.

Fig.~\ref{Fig2}(b) displays a band dispersion along the $k_z$ axis (along the A-$\Gamma$-A line), obtained by varying the photon energy of the incident light source (see Methods). 
There are three dispersive bands visible within the energy range shown, all of which are of Se $p$ character~\cite{eknapakul_direct_2018}. At $E-E_{\text{F}} \approx $ -1.2 and -1.6~eV, a pair of less-dispersive bands are visible. 
We assign these to be the pair of spin-orbit split Se $p_{x,y}$-derived bands. Their limited dispersion along $k_z$ originates from the disparity in the hopping strengths of $p_x$ and $p_y$ orbitals in the $x$-$y$ plane compared to that across the van der Waals gap. It is this quasi two-dimensionality that is thought to enable a hidden spin polarisation of these bands, with their wavefunctions being sufficiently localised to one half of an X-M-X unit. 
Also visible in Fig.~\ref{Fig2}(b) is a strongly dispersing band spanning from the valence band maximum at $E-E_{\text{F}}\approx$-1.1~eV down to $\approx$-3.8~eV. This band crosses through the $p_{x,y}$-derived states in the vicinity of the VBM. We attribute this band to be of Se $p_z$ derivation, with enhanced hopping along the $c$-axis deriving from the extended spatial extent of $p_z$ orbitals into the van der Waals gap. The remaining two-dimensional band visible at $E-E_{\text{F}} \approx $ -4.1~eV is not of relevance here.

As a point of interest, we note that this crossing pattern between three-dimensional chalcogen $p_z$-derived bands and quasi two-dimensional chalcogen $p_{x,y}$ bands along this rotationally $C_{3v}$-symmetric $\Gamma$-A line is known to generically generate highly tunable so-called  `topological ladders' across the TMD family~\cite{bahramy_ubiquitous_2018, ClarkGeneral2019, xiao_manipulation_2017}.  
In short, due to the influence of trigonal crystal field on the chalcogen $p$-orbital manifold, the crossing between each $p_z$-derived band with each pair of $p_{x,y}$-derived bands produces a single symmetry-protected bulk crossing point (or so-called bulk Dirac point), and a hybridisation gap (HG) which is often topologically non-trivial~\cite{yang_classification_2014}. In Fig.~\ref{Fig2}(c), we enhance the region of the band dispersion where the crossing pattern occurs {and overlay orbitally-projected density functional theory (DFT) calculations. Each band is labelled with the corresponding space group representation.} A hybridisation gap between the $p_z$-derived band (labelled $R_{4}$) and the lower energy $p_{x,y}$ band ($R_{4'}$) is clear, although there is little evidence of a two-dimensional state occupying the gap. This suggests a topologically trivial classification for the hybridisation gap. It follows that a crossing between the $p_z$-derived $R_4$ and the top-most $p_{x,y}$-derived state ($R_{5,6}$) would generate bulk Dirac points symmetrically located about $\Gamma$. However, due to a suppression in the photoemission matrix element of the $R_{5,6}$ band in the vicinity of the $\Gamma$ point, we cannot rule out that it avoids $R_4$ entirely. In other words, the presence of bulk Dirac points in this compound is dependent on whether or not the valence band top is of $p_z$ or $p_{x,y}$ character. 
{Our DFT calculations, which account for electron correlation effects (see Methods), predict that this crossing point does occur, with a 20~meV gap separating the $p_z$-derived VBM and the $p_{x,y}$-derived band below at the $\Gamma$ point. The valence band top of HfSe$_2$ can therefore be described as a pair of closely-spaced type-II bulk Dirac cones of the same origin as those widely studied in the (semi-)metallic group X TMDs, recently found to have applications in THz optoelectronics~\cite{zhang_high_2021}.}

In Figure~\ref{Fig2}(d-g), we turn to the in-plane band dispersions of these Se $p$-derived states along the $\overline{\mbox{K}}$-$\overline{\Gamma}$-$\overline{\mbox{K}}$ direction of the surface BZ for various photon energies corresponding to $k_z$ planes between A and $\Gamma$. The significant in-plane dispersion of the relatively sharp $p_{x,y}$-derived bands is clear, sitting against a background of diffuse spectral weight forming umbrella-like dispersions down to approximately $E-E_{\text{F}}$ = -3.8eV in each case. While a chosen photon energy will selectively excite a specific $k_z$ plane, due to the surface sensitivity of our photoemission experiments, the envelope of $k_z$ integration can span over several BZs~\cite{Dam2004,waditi_angle_2006}. The severity of this effect correlates to the extent of the band dispersion along $k_z$. The $k_z$-broadened spectral weight in Fig.~\ref{Fig2}(d-g) is therefore predominantly of $p_z$ character. In Supplemental Fig.~S1~\cite{Sup}, we further explore the symmetry of the pair of $p_{x,y}$-derived bands as they disperse in-plane as function of photon energy. While these bands are very two-dimensional relative to the $p_z$-derived band, there is a periodic evolution in how these bands disperse away from the $\overline{\Gamma}$ point as a function of $k_z$, providing further confirmation that these are indeed bulk states.

\

\noindent{\textbf{Apparent time-reversal symmetry breaking in HfSe$_2$}}

Next, we turn to the spin-polarisation of the bands discussed in the previous section. There are {several} possible sources of spin-polarisation here. The 1T-structured TMDs are known to have several surface states both topologically trivial and non-trivial with complex band dispersions due to the influence of the dense $k_z$-projected bulk band manifold~\cite{ClarkGeneral2019}. Any such states are permitted to be spin-polarised due to the inversion asymmetry offered by the surface potential step. These should be Rashba split with a predominantly in-plane momentum-locked chiral spin texture. 
We note that there is little evidence of surface states in the data presented in Fig.~\ref{Fig2}, and so this origin is unlikely. The other source is a spin-polarisation of bulk bands caused by the hidden-spin mechanism discussed in Fig.~\ref{Fig1}. Although unobserved to date, this too should produce a predominantly in-plane chiral spin texture, and is expected to be most prominent in the $p_{x,y}$-derived valence bands which are more likely to have sufficiently localised wavefunctions, as evidenced by their relative two-dimensionality (Fig.~\ref{Fig2}). {Finally, we note that the superposition of incoming and reflected wavefunctions from the surface potential step in elemental metals has been demonstrated to result in observable Rashba-type spin polarisations of bulk bands when probed by ARPES, with an analogous mechanism possibly applicable to other systems~\cite{kimura_strong_2010,krasovskii_rashba_2011}. }
We stress that measured spin-polarisations deriving from any of these origins should be strictly time-reversal symmetric.

\begin{figure}[t]
	\centering
	\includegraphics[width=\columnwidth]{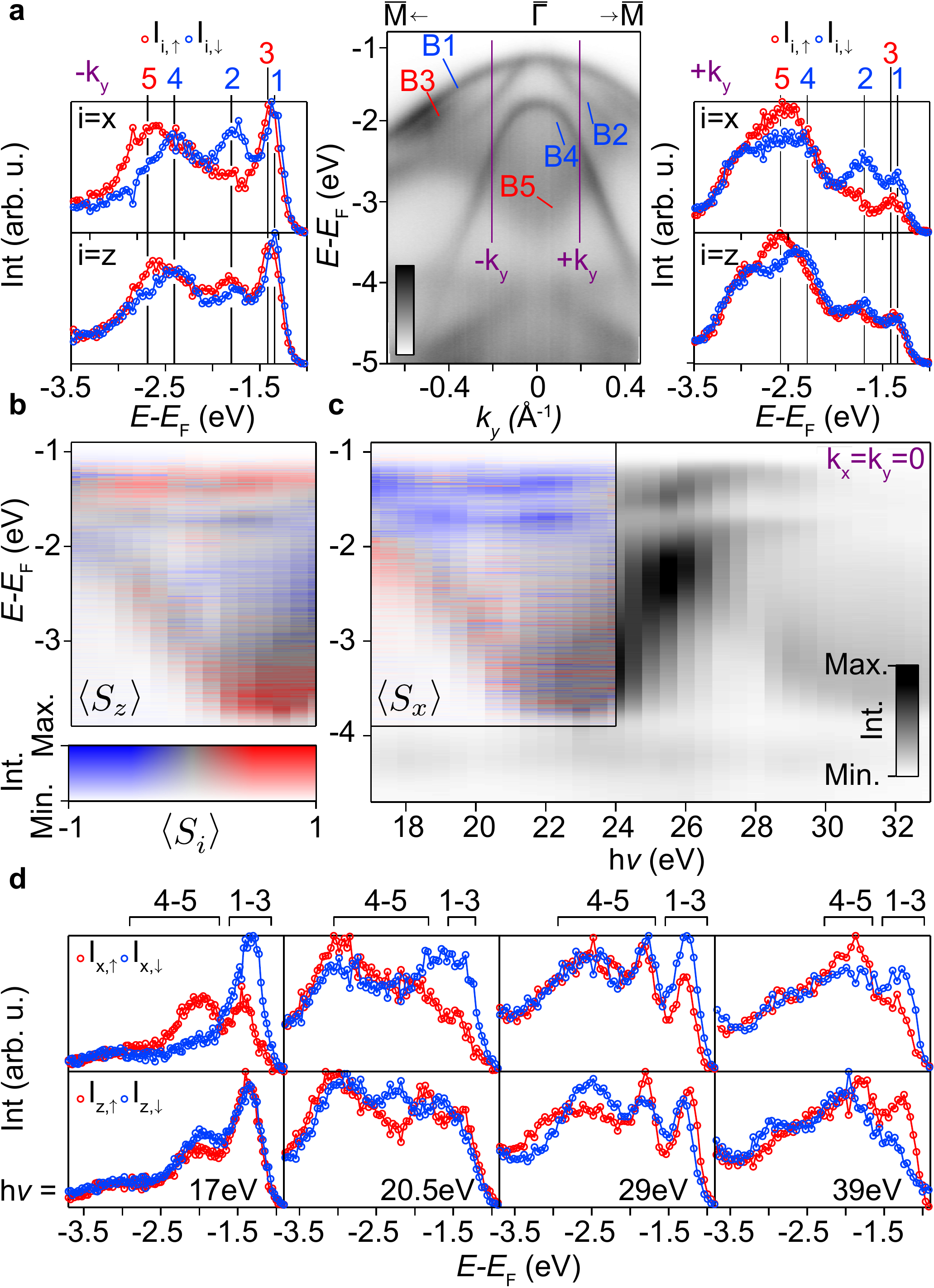}
	\caption{\label{Fig3} \textbf{Apparent TRS breaking in HfSe$_2$} (a) spin-EDCs { (h$\nu$=29~eV)} for the $x$ ($\parallel k_x$) and $z$ components of spin for two $k$ points indicated in the central ARPES spectrum taken along the $\overline{\Gamma}$-$\overline{\mbox{M}}$ direction. (b-c) Extracted spin-polarisation for spin EDCs taken at $k_x=k_y=0$ as a function of $k_z$ for the (c) $x$ and (b) $z$ components. A spin-integrated photon energy dependence over the same $E$-$h\nu$ range is used as an intensity filter (note the two-dimensional color bar). Five pairs spin-resolved EDCs and eight spin-integrated EDCs were used to generate each spin-resolved image. (d) Select spin-EDCs for the $x$ and $z$ components for 17, 20.5, 29, and 39eV incident photons, from left to right. The analyser slit is defined as parallel to the $k_x$ direction for all measurements here. }
\end{figure}

In Fig.~\ref{Fig3}(a), spin-resolved energy distribution curves (spin-EDCs) are shown on each side of the time-reversal invariant momentum $\overline{\Gamma}$ along the $\overline{\mbox{M}}$-$\overline{\Gamma}$-$\overline{\mbox{M}}$ ($k_y$) direction, alongside a spin-integrated band dispersion taken with 29~eV photons corresponding to a $k_z$ plane where the $p_z$ band is at the shallow binding energy turning point (Fig.~\ref{Fig2}(b)). {Accounting only for the exponential attenuation of the emitted photoelectrons, for this photon energy we estimate that 45\% and 64\% of the photoemission intensity will originate from the top sub-layer of the unit cell, and from sub-layers equivalent to the top sub-layer, respectively~\cite{ seah_surface_1979}. This is sufficiently sensitive to probe any hidden spin polarisation, even when assuming purely destructive interference between the two structure types. We note, however, that the interference pattern is likely far more complex~\cite{riley_direct_2014}.} There are five bands labelled in the central ARPES image. The states B1 and B2 together form the valence band top of HfSe$_2$ for this photon energy. B1 is the $p_z$-derived band, originating from the shallow-energy turning point along $k_z$, which is less dispersive in-plane. B2 is the shallower binding energy $p_{x,y}$-derived state from previous discussions, forming a narrower parabola in the vicinity of the VBM.  B4, also very dispersive in plane, is the second $p_{x,y}$-derived state from previous discussions.  The high binding energy apex of the $p_z$-character, $k_z$-broadened spectral weight (originating from the turning point of the $p_z$-derived band in $k_z$ near the A point) is labelled as B5.
A final band, B3, visible only at higher $|k_y|$ values, appears to form a pair with either B1 or B2, separating only away from $\overline{\Gamma}$. This description is consistent with a splitting of both $p_{x,y}$-derived states shown in earlier DFT studies~\cite{eknapakul_direct_2018}, {but we note that are other possible origins: A localised turning point or plateau in the $k_z$ dispersion of B1 or B2 could give the impression of a separate, distinct band in $k_z$-broadened spectra. We also note that a single-branch state that disperses exactly through the $k_z$-projected type-II Dirac nodes was observed by ARPES in both PdTe$_2$ and PtSe$_2$, and was not replicated by the accompanying band structure calculations~\cite{bahramy_ubiquitous_2018, ClarkGeneral2019}. Although not well understood, one may expect a similar state here given the common origin of the Dirac nodes.}

Turning attention to the spin-EDCs, it becomes clear that all five of these states are spin-polarised, with signal in each of the measured  in-plane chiral ($x$) and out-of-plane ($z$) channels. B1, B2 and B4 have an opposite, predominantly in-plane, spin-polarisation to B3. The diffuse $p_z$ weight from the higher binding energy turning point in $k_z$, B5, carries a pronounced polarisation, opposite to that of the shallower energy turning point, B1. This suggests that the bulk $p_z$ band carries a spin-polarisation switching from $\Gamma$ to A. This final observation is particularly surprising, suggesting that $p_z$ orbitals can also carry a hidden spin-texture despite their higher $c$-axis delocalisation relative to the $p_{x,y}$ orbitals. {It is likely, therefore, that proximity to the more two-dimensional states  energetically positioned near the extrema of the $p_z$-band $k_z$ dispersion~\cite{eknapakul_direct_2018} produces a hybridised state of mixed character, with a wavefunction sufficiently localised to carry a hidden-spin.}

A second set of spin-EDCs is displayed for a $k_y$ point on the other side of  $\overline{\Gamma}$. Astonishingly,  the direction of the spin-polarisation of all states remains unchanged when transitioning from $-k_y$ to $+k_y$, strongly suggesting a TRS-breaking spin texture for each state. To further investigate the apparent TRS breaking, Fig.~\ref{Fig3}(b-d) show spin-EDCs taken at the $\overline{\Gamma}$ point ($k_x=k_y=0$) for a series of photon energies covering approximately one $\Gamma$-A line in $k_z$.
As both A and $\Gamma$ are TRIM points, all bands along the entire A-$\Gamma$-A line should be entirely spin degenerate in the absence of magnetism. Fig.~\ref{Fig3}(b) and (c) show measured spin-polarisations in the $z$ and $x$ channels respectively as a function of photon energy, with the spin-integrated ARPES image overlaid as an intensity filter (note the two-dimensional colour bar). The spin-EDCs in Fig.~\ref{Fig3}(d) are select EDCs from these same datasets along with higher energy EDCs taken with $h\nu=$29 and 39~eV. Despite conventional wisdom stipulating that there should be strictly no spin-polarisation at $\overline{\Gamma}$ for any choice of $k_z$, these data unambiguously demonstrate that the bands discussed previously retain a strong non-zero spin-polarisation both at the $\overline{\Gamma}$ point and for all $k_z$. We believe this is the first such observation of $\overline{\Gamma}$ point spin-polarisation in a non-magnetic system. We note that the persistence of spin polarisations that remain largely unchanged for a large range of incident photon energies precludes a final state origin~\cite{sanchez-barriga_photoemission_2014}.

While these observations are seemingly at odds with the fundamental symmetries possessed by this compound, it is possible to reconcile the experimentally-obtained spin texture with a hidden-spin mechanism that is entirely respectful of TRS. The solution has two distinct parts, requiring both the consideration of individual $p$-orbital contributions to the overall spin texture, and the role played by orbital-selective photoemission matrix elements.

\

{
\noindent{\textbf{Local spin-orbital magnetisation in HfSe$_2$ subcells}}

In attempt to uncover the origin of the apparent TRS-breaking shown in the previous section, in Fig.~\ref{Fig3b}(a), we display the spin-resolved valence electronic structure for HfSe$_2$, as determined by density functional theory (DFT) (See Methods) for HfSe$_2$ along the M-$\Gamma$-M direction. The chiral component of spin, parallel to $k_y$ here, is projected onto the top (1) and bottom (2) Se lattice sites, verifying the presence of a hidden spin mechanism in this compound.

The electronic structure of the valence bands itself arises from the interplay of the $C_{3v}$ crystal field and spin-orbit interaction (SOI). The former acts on both B and AB groups individually, splitting each into two sub-manifolds $\{p_x,p_y\}$ and $p_z$. Mixed by SOI, they further reduce to three doubly degenerate $|J,\pm m_j\rangle$ branches with $J=3/2$ and 1/2, and $1/2 \le m_j \le J$. Of highest relevance here is the AB group, composed of the three shallowest-binding energy bands in Fig.~\ref{Fig3b}(a). The Se sublayer-localised chiral spin textures demonstrate the symmetry-imposed interplay between the underlying lattice with the spin and orbital degrees of freedom of the occupying electrons. The two sites have opposite chiralities to respect the overall TRS and global IS of the system, consistent with the simplified Rashba-model discussed in Fig.~\ref{Fig1}(d).

In-line with previous predictions~\cite{cheng_hidden_2018}, these calculations show, therefore, that a sufficiently sensitive probe of the electronic structure would indeed measure a non-zero spin-polarisation of the near-$E_{\text{F}}$ Se $p$-derived bands, exactly switching from Se1 to Se2, but clearly do \textit{not} predict symmetric spin-polarisations of any state about $\overline{\Gamma}$ like those observed experimentally in Fig.~\ref{Fig3}. The behavior of the $p_z$-derived state is well reproduced, however, exhibiting a finite spin-polarisation despite the expectation of a more delocalised wavefunction along the $c$-axis relative to the $p_{x,y}$-derived bands. This is further explored in Fig.~\ref{Fig3b}(b), where the $k_z$-projected band dispersion shows how a hidden spin-polarisation of all bands is present for all $k_z$, with a reversing polarisation of the $p_z$-derived between the turning points of its $k_z$-dispersion, and with an enhanced polarisation magnitude relative to the intermediate $k_z$ planes. This behavior highlights the significant role played by spin-orbital mixing within the Se $p$-orbital manifold.

\begin{figure}[t]
	\centering
	\includegraphics[width=\columnwidth]{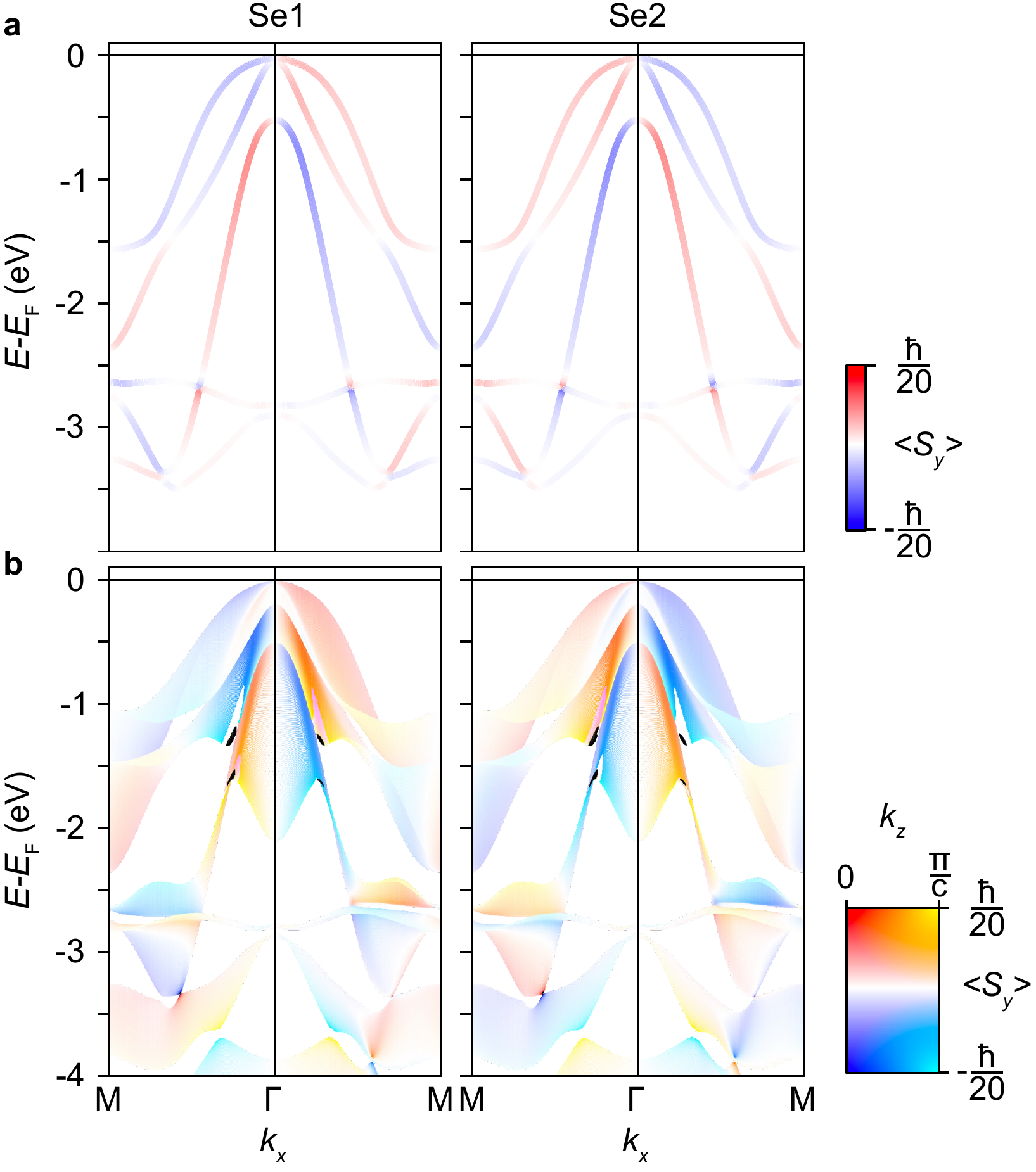}
	\caption{\label{Fig3b} \textbf{Se-site projection of the electronic structure from Se $p$-orbitals.} (a) The combined $p$-orbital contribution to the spin-texture of HfSe$_2$ is shown, projected onto the top (Se1, left) and bottom (Se2, right) layer. (b) Equivalent calculations to those in (a), now including a full $k_z$ projection for better comparison to experiment (see two-dimensional colour bar).}
\end{figure}

Now let us consider the distinct individual $p$-orbital contributions to the hidden spin-polarisation shown in Fig.~\ref{Fig3} and Fig.~\ref{Fig3b}. This distinction is significant for two reasons. Firstly,} the driving force behind the hidden spin mechanism in the 1T class of TMDs is the $c$-axis aligned net electric fields within each $C_{3v}$-symmetric HfSe$_6$ octahedron. The orthogonal Se $p_{x,y,z}$ orbitals within each Se sublayer experience the trigonal field disparately, and so this asymmetry should be reflected by the individual $p$-orbital contributions to the spin texture of the locally Rashba-split $p$-derived bands. Inequivalent orbital contributions to a global spin texture, or `coupled spin-orbital textures', have been previously discussed in the context of the `giant' Rashba-split semiconductor BiTeI and the topological insulator Bi$_2$Se$_3$~\cite{bawden_hierarchical_2014,zeljkovic_mapping_2014,lin_orbital_2018,beaulieu_revealing_2020}. Secondly, while the total $p$-orbital contribution to the $\overline{\Gamma}$ point would still be expected to sum to zero in accordance with TRS, orbitals are not excited equally in the photo-excitation process. {Indeed, if one considers any single band in Fig.~\ref{Fig3b} without its energy-degenerate partner, a spin-orbital magnetisation is uncovered, with the $p_x$ and $p_y$ contributions to the spin polarisation at the $\overline{\Gamma}$ point finite and exactly opposite. An orbital-selective probe of electronic structure, applicable to the present experiment as explained in the next section, could then selectively probe the spin-texture partially and thus detect a net spin polarisation at $\overline{\Gamma}$. However, the energy degenerate partner behaves exactly oppositely, negating this effect. 

\begin{figure}[t]
	\centering
	\includegraphics[width=\columnwidth]{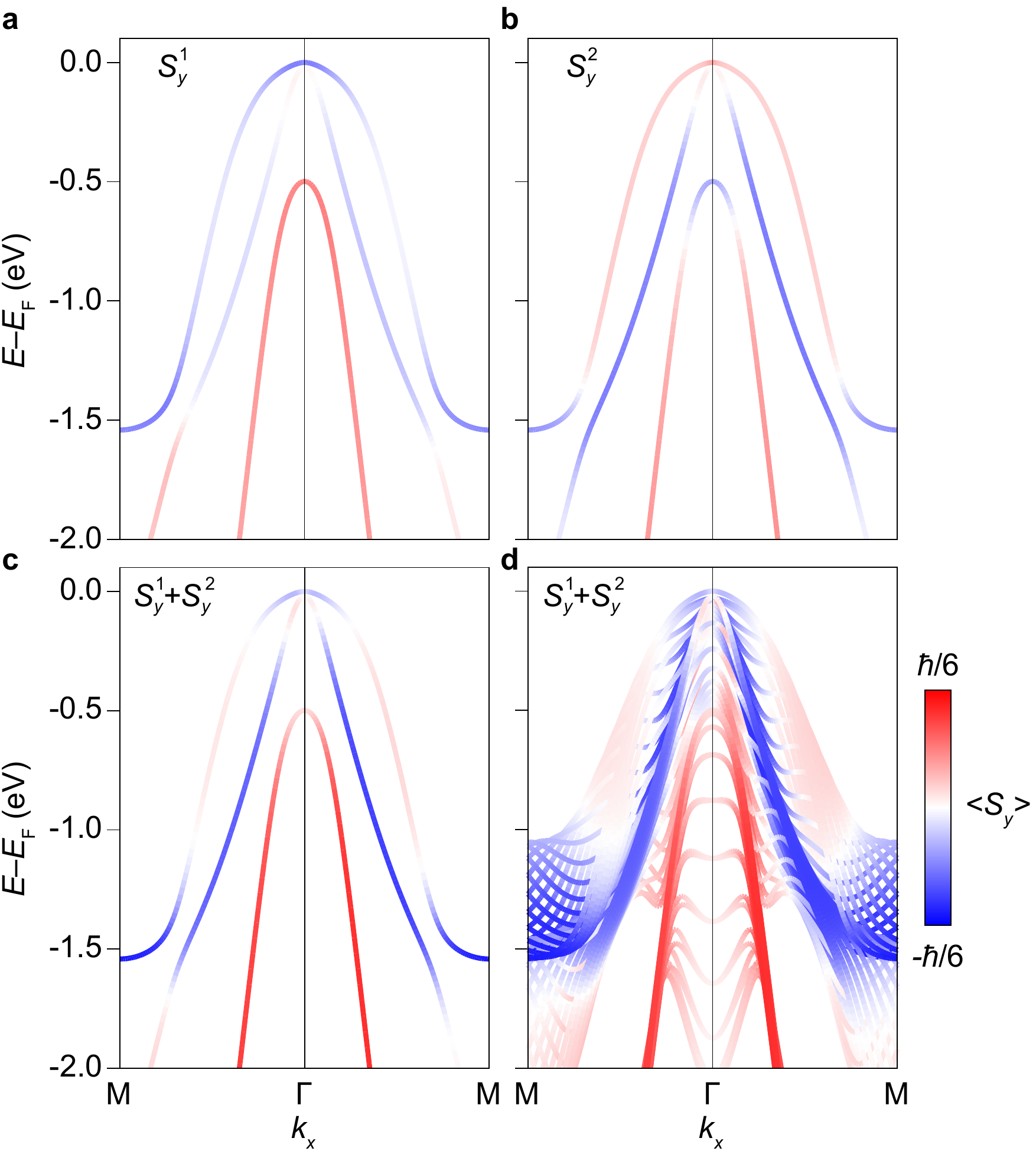}
	\caption{\label{Fig3c} \textbf{Contributions from $p_{x,y}$-$p_z$ overlap in the vicinity of Se1} (a-b) Spin-resolved electronic band structure contribution from $S_y^1$ and $S_y^2$ overlap integrals projected onto Se1 along M-$\Gamma$-M. (c-d) The combined contribution $S_{y}^{1}+S_{y}^{2}$ is shown for both the $k_z$=0 plane (c) and for all $k_z$ (d). }
\end{figure}

Instead, motivated by the substantial contribution of the $p_z$-derived band in both Fig.~\ref{Fig3} and~\ref{Fig3b}, we further explore the role of spin-orbital mixing terms between constituent $p$-orbitals. To facilitate this, we form a 12-band tight-binding model from the $|J,m_j\rangle$ states, \mbox{$\psi_v=\sum_{J,m_j} \alpha_{J,m_J} |J,m_j\rangle$}. This framework, detailed in full in Supplementary Note 2~\cite{Sup}, enables the computation of individual Se $p$-orbital  contributions to the three component $i=\{x,y,z\}$ spin-polarisation localised at each Se site, $a$, using  $S_{i,a}=\langle \psi_{v,a}| \sigma_i | \psi_{v,a}\rangle$, where $\sigma_i$ are Pauli spin matrices.

Restricting the problem to the vicinity of a single Se layer, we find four overlap integrals $\langle J, m_J | \sigma_i |J', m_{J'} \rangle$ with non-vanishing values at the $\Gamma$ point. These terms obey $\Delta J = J-J' =0$ or 1 and $\Delta m_J = m_J - m_{J'} = 2$  All the other overlap integrals vanish at the $\Gamma$ point due to TRS (see Supplementary Note 2 for a full derivation~\cite{Sup}). The resulting spin polarisation from these four terms undergoes a sign reversal when switching between Se sites, ensuring full compliance to TRS and IS. A careful inspection of the above $\Delta J$ and $\Delta m_J$ constraints reveals that only the spin-orbital terms mixing $\{p_x, p_y\}$ and $p_z$ sub-manifolds can contribute to such a non-vanishing spin polarisation at a time-reversal invariant momentum. This finding further signifies the critical role of $p_z$ orbitals in the electronic and spin structures of the low energy bands in HfSe$_2$. It should be noted that all overlap integrals can hold finite contributions away from the $\Gamma$ point, altogether producing the chiral spin textures previously discussed in Fig.~\ref{Fig3b}. 

Fig.~\ref{Fig3c}(a) and (b) show the spin-projected band structure stemming from these $\Delta J=0$ and 1 terms, denoted as $S_y^1$ and $S_y^2$, respectively, for the chiral ($\sigma_y$) component at Se1. These exhibit symmetric spin-polarisations about $k=0$ as for the experimentally determined spin texture. This suggests that orbital mixing of in-plane and out-of-plane orbitals is the driving force behind the the experimentally-observed symmetric spin-polarisations. It is noteworthy that these two in-equivalent contributors are not opposite, with their sum and $k_z$-projection shown in Fig.~\ref{Fig3c}(c-d). This demonstrates that a small net spin-polarised contribution to the BZ centre detectable even to a probe of electronic structure blind to orbital type. The two non-zero contributions to the spin-texture can be tuned between, however, and therefore enhanced, with a linearly polarised light source. This provides a convenient way to further verify this description, and explore routes towards the tunability of this effect for applications. }

\

\noindent{\textbf{Disentangling local spin-orbital textures at $\overline{\Gamma}$}}

To experimentally confirm that the hidden spin-orbital textures, {driven by the spin-orbit mixing between $p_z$ and in-plane $p_{x,y}$ orbitals}, are indeed the source of the observed spin-polarisation, {we can tune the orbital selectivity of the experimental probe and test that the measured spin polarisation responds accordingly.} The photoemission matrix element for a transition between an initial ($i$) and final ($f$) is given as $|M_{f,i}^{{k}}|^2 \propto |\langle \phi_f^k|\vec{A} \cdot \vec{p} |\phi_i^{{k}}\rangle|^2$, where $\phi_{i,f}^{{k}}$ are initial and final state wavefunctions for an electronic band at $k$, and $\vec{A}$ and $\vec{p}$ are the  vector potential and the momentum operator respectively~\cite{Dam2004}. This integral is required to be of overall even parity to be non-zero. Assuming an even parity final state, this integral is non-zero for initial state $p_x$ and $p_z$ orbitals with a $p$-polarised light source. Switching to $s$-polarised light would probe $p_y$ orbitals in isolation.
Therefore for either choice of linear light polarisation, the ground state spin-orbital texture is only partially probed, aiding the observation of apparent time-reversal symmetry breaking spin textures.

\begin{figure}[t]
	\centering
	\includegraphics[width=\columnwidth]{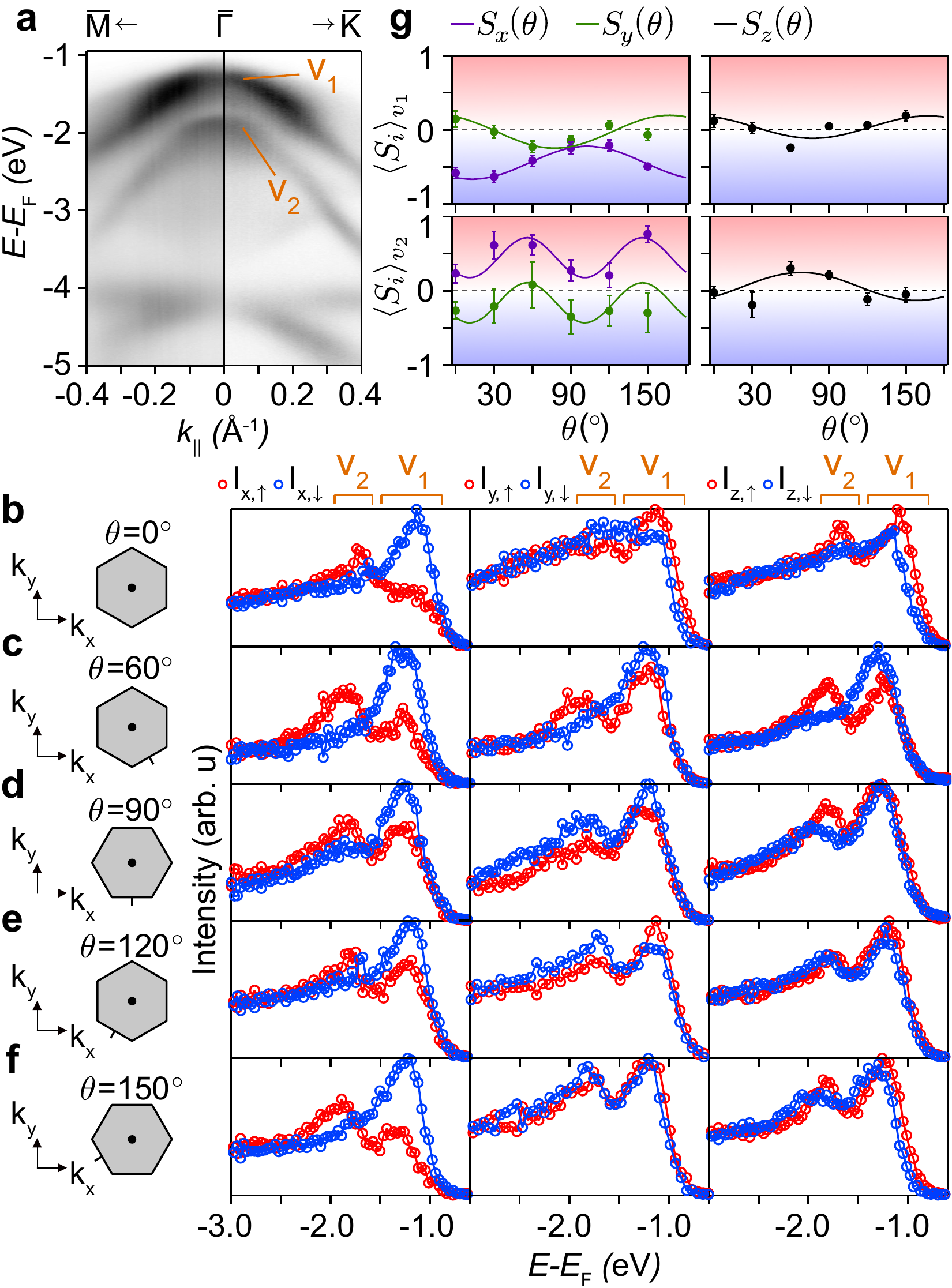}
	\caption{\label{Fig4} \textbf{Accessing the full spin-texture of the VBM} (a) Dispersion along $\overline{\mbox{M}}-\overline{\Gamma}-\overline{\mbox{K}}$ taken with 17~eV photons. The collection of bands $v_1$ and $v_2$ are labelled. (b-f) A subset of spin-EDCs used for polarisation extraction in (f). The in-plane sample rotation, $\theta$, is indicated in each case. $x$, $y$ and $z$ components are shown in the left, middle, and right- hand columns respectively. (g) Extracted $\overline{\Gamma}$-point spin-polarisations along the $x$, $y$, and $z$ directions as a function of sample rotation in the $x$-$y$ plane ($\theta$) for bands $v_1$ (top) and $v_2$ (bottom). Solid lines are functions satisfying the TRS-preserving model discussed in Supplementary Note 3~\cite{Sup}. The analyser slit is defined as parallel to the $k_x$ direction for all measurements here. 
	}
\end{figure}

In Fig.~\ref{Fig4}, we continuously rotate our sample in the $x$-$y$ plane while keeping a fixed $p$-polarised light source. A rotation of $\pi/2$  switches the parity of the photoemisison matrix elements for $p_x$ and $p_y$ orbitals, while leaving that for $p_z$ orbitals unchanged. By performing a series of spin-EDCs at $\overline{\Gamma}$ as a function of the in-plane sample azimuthal angle $\theta$ with a spin-detector capable of probing the spin-polarisation in all three-directions, it is therefore possible to evaluate the spin-polarisation contribution from $p_x$, $p_y$ and $p_z$ orbitals in full, {here allowing for the tuning the relative weight of the contributions from $S_y^1$ and $S_y^2$ overlap integrals due to their in-equivalent in-plane orbital compositions.}

Figure~\ref{Fig4} provides an overview of the $\overline{\Gamma}$ point spin polarisation as a function of $\theta$ for 17~eV photons, where the spin polarisation is particularly clear, as also seen in Fig~\ref{Fig3}(d)). Fig.~\ref{Fig4}(a) shows the valence band structure along a $\overline{\mbox{K}}-\overline{\Gamma}-\overline{\mbox{M}}$ path. There are two bands visible at $\overline{\Gamma}$ with energies $E-E_{\text{F}}=$ -1.3 and -1.7~eV. By cross-referencing to the spectra in  Fig.~\ref{Fig3}, it can be determined that, at $\overline{\Gamma}$, the top band labelled $v_1$ is an amalgamation of up to three (doubly-degenerate) bands, previously labelled as B1, B2 and B3. The lower band, labelled $v_2$ is likely predominantly made up of band B4 discussed previously. Fig.~\ref{Fig4}(b) show a series of three-component spin-EDCs measured at 30 degree increments in $\theta$. For clarity, we fix the $k_x$ and $k_y$ plane to be parallel and perpendicular to the initial  $\overline{\mbox{M}}-\overline{\Gamma}-\overline{\mbox{M}}$ orientation at $\theta=0$, respectively. The spin components $S_x$, $S_y$ and $S_z$ are the measured spin signals in this fixed laboratory reference frame. 

A subset of the resulting series of spin-EDCs are shown in Fig.~\ref{Fig4}(b-f). 
There is clear periodicity to the measured spin polarisation in each channel. The $S_x$ projection for $nu_1$ is dominant for all angles, and the sign of the $S_x$ component is independent of $\theta$ for both $nu_1$ and $nu_2$.  This is not the case for either the $S_y$ or $S_z$ channels, with the net polarisation within $v_1$ switching sign periodically in both channels. For $v_2$, the $S_z$ spin polarisation magnitude is strongly $\theta$-dependent, and has a reversing signal in $S_y$. {These qualitative observations already show that photoemission matrix elements strongly influence the measured spin-polarisation at $\overline{\Gamma}$, independently suggesting a spin-orbital locking of the measured spin-signal~\cite{Bawden2015}.}

To better visualise the changes imposed by the sample rotation, we extract the spin-polarisation from the peaks corresponding to $v_1$ and $v_2$ as a function of $\theta$ (see Methods). The result of this analysis is displayed in Fig.~\ref{Fig4}(g). The overlaid solid lines are periodic functions that satisfy a simplified model that assumes that TRS is in-tact (see Supplementary Note 3~\cite{Sup}). Specifically, it assumes that the $p_x$, $p_y$ and $p_z$ contributions to the $\overline{\Gamma}$-point spin polarisation exactly cancel out, and therefore that the observation of non-zero spin is only possible with an incomplete probing of the full set of $p$-orbitals, {although as shown in the previous section, there is a route towards finite spin-polarisation even with an unpolarised light source due to a finite $S_y^1+S_y^2$ sum}. The second constraint is more robust, however, restricting the periodicity to the measured spin-polarisation functions to be strictly of $\pi/n$, where $n\in \mathbb{Z}$. Satisfying this second constraint entirely removes the possibility of spin vectors of a magnetic origin, as the photoemission orbital-selective matrix elements in 0 and $\pi$ rotated geometries are equivalent, and so an otherwise equivalent experiments should yield equivalent results.  A real (in-plane) magnetic spin vector would always switch sign with a $\pi$ rotation. 

All functions do satisify this constraint for both the $v_1$ and $v_2$ peaks, with the $S_x$ and $S_y$ signals sharing the same periodicity. For the higher (lower) binding energy state, $v_2$ ($v_1$), the spin-polarisations in the $x$ and $y$ channels are well fit to a pair of functions that are $\pi/2$ ($\pi$) periodic. The measured $S_z$ polarisation is $\pi$ periodic for both $v_1$ and $v_2$. We note that the measured spin-polarisation functions only satisfy the simplified TRS-conserving model if there is a $p_z$ contribution to the spin-polarisaiton. Due to the effective double-counting of the $p_z$ contribution as the sample is rotated, the spin-polarisation functions in the $S_x$ and $S_y$ channels do not have to be exactly opposite, instead summing to a periodic function with limits defined by the $p_z$ contribution to the in-plane channels. This observation is entirely consistent with the crucial role played by $p_z$ orbitals uncovered in the previous section. The measured spin-polarisation at $\overline{\Gamma}$ is thus consistent with a time-reversal symmetric spin-orbital texture driven by the mixing of $p_{x,y}$ and $p_z$ orbitals in the ground state.

Taken altogether, we conclude that the presence of a non-zero spin-polarisation at the BZ centre of HfSe$_2$ is a direct consequence of a coupled hidden spin-orbital texture. The adheration of the measured spin-polarisations to a model in which time-reversal symmetry is not broken again confirms that the hidden coupled spin-orbital texture observed here is a ground state property of bulk HfSe$_2$. Note that an exactly equivalent conclusion would be drawn when using $s-$polarised light, but with measured spin polarisation enitrely composed of the missing half of the $p_{x,y}$ contribution.

\

\noindent{\textbf{Application for transient spin-current generation}}

To finalise, we briefly discuss potential for this type of $\Gamma$ point orbital-selectivity in the context of applications in spin-polarised current generation. As demonstrated in Fig.~\ref{Fig3c} and~\ref{Fig4} and discussed in the previous section, the sign of the spin-polarised signal from the valence band maximum of HfSe$_2$ is dependent on the subset of the $p$-orbital contributions probed, thus opening opportunities to  populate the unoccupied conduction bands with electrons belonging to a chosen spin-species, with linearly polarised, fixed energy light sources. While it is important to ensure that the mean free path of photo-excited elections remains  small enough to be sensitive to the hidden-spin derived spin-polarisation, there is some scope to tune final states with the choice of photon energy.

These spin-selective excitations are different to those possible in the monolayer 1H-TMDs, where the transient spin polarisation of  the conduction band minimum following excitation is dependent both on the helicity and the energy of a circularly-polarised light source~\cite{mak_control_2012,xaio_coupled_2012}.
Moreover, any rotational domains in the 1H and 2H-TMDs could randomize the spin signal from the oppositely polarised K and K' points, requiring high-quality single crystals with a known orientation. For HfSe$_2$, we demonstrated in Fig.~\ref{Fig4} that changing the orientation of the crystal does not switch the sign of the dominant spin channel, greatly reducing any hindrance from the presence of multiple rotational domains. As a result, poly-crystalline samples may also yield similar net-polarisations to those seen here when using linear polarised light sources, potentially reducing barriers to commercialisation. 

We expect that the new physics observed here is general to the other semiconducting 1T structured TMD without a significant transition metal $d$-derived valence band contribution, although likely less pronounced than the present case due to a reduced dipolar field strength within the IT-unit. For the (semi-)metallic systems (e.g. (Ir, Ni, Pt, Pd)(Se,Te)$_2$) it is likely that screening effects suppress the effective crystal field splitting within chalcogen sites key to the phenomena observed here. 

\section{Conclusion}

We have shown that the $\overline{\Gamma}$-centred, selenium $p$-orbital derived valence bands of 1T-HfSe$_2$ exhibit hidden ground state spin-polarisation owing to the strong out-of-plane dipole present within each half Se-Hf-Se sublayer of the unit cell. This hidden-spin polarisation manifests as a coupled spin-orbital texture wherein $p_x$, $p_y$ and $p_z$ orbitals contribute differently to the Rashba-split states, and therefore a net spin-polarisation at the $\overline{\Gamma}$ point can be measured and enhanced when selectively probing a subset of these orbitals, despite the presence of time-reversal symmetry. These findings offer a new route towards the generation of spin-polarised carriers from poly-crystalline 1T-structured transition metal dichalcogenides by using a fixed energy, linearly polarised light source, revealing possibilities to exploit effective net spin polarisations in non-magnetic systems more widely.

\phantom{xxxx}




\

\noindent {\bf Acknowledgements}

\noindent{The authors thank Lewis Bawden for useful discussions.  We acknowledgde financial support from the Impuls- und Vernetzungsfonds der Helmholtz-Gemeinschaft under grant No. HRSF-0067.} 

\noindent{This version of the article has been accepted for publication, after peer review and is subject to Springer Nature’s AM terms of use, but is not the Version of Record and does not reflect post-acceptance improvements, or any corrections. The Version of Record is available online at: https://doi.org/10.1038/s41467-022-31539-2}

\

\noindent {\bf Author Contributions}
\noindent {O. J. C. performed the experiment, analysed the experimental data, developed the model describing the periodicity of the measured spin-polarisation functions and wrote the manuscript with input from all co-authors. J. S.-B. aided with data collection and interpretation, and is responsible for the U125-PGM beamline and its spin-ARPES endstation. Following the initial Reviewer comments, O. D. and M. S. B. performed density functional theory calculations and developed the tight-binding model discussed in the Supplementary Information. O. J. C. was responsible for the overall project planning and direction.}

\

\noindent {\bf Data Availability}

\noindent {Data are available from the authors upon reasonable request.}

\phantom{xxxx}

\phantom{xxxx}


\clearpage

\section{Methods}

\noindent{\textbf{Experimental:}}
ARPES measurements were carried out using $p$-polarised synchrotron light of energies between 17 and 64~eV at the U125-PGM beamline of BESSY-II in Helmholtz-Zentrum Berlin.
Photoelectrons were detected with a Scienta R4000 analyzer at the Spin-ARPES endstation, and the base pressure of the setup was better than $\sim$1$\times$10$^{-10}$ mbar.  High-quality HfSe$_2$ single crystals were cleaved \textit{in situ} using a standard top post method at temperatures of 40-50~K. The data presented are from three samples, all of which showed qualitatively the same behavior, including in the absolute directions of spin-polarisation. The spin data in Fig.~\ref{Fig3}(a), Fig.~\ref{Fig3}(b-d) and Fig.~\ref{Fig4} originate from samples 1, 2 and 3 respectively.  No sample was magnetised.

To determine the HfSe$_2$ $k_z$ dispersion from photon-energy-dependent ARPES, we employed a free electron final state model
\begin{equation}
k_z=\sqrt{\frac{2m_e}{\hbar^2}} (V_0+E_k  \cos^2\theta )^{1/2}
\end{equation}
where $\theta$ is the in-plane emission angle and $V_0$ is the inner potential. We find best agreement to the periodicity of the dataset when taking an inner potential of 10.5 eV for a $c$-axis lattice constant of 6.160 \AA.

Spin resolution was achieved using a Mott-type spin polarimeter operated at 25 kV and capable of detecting all three components of the spin polarisation. Spin-resolved energy distribution curves (spin-EDCs) were determined according to 

\begin{equation}
	I_i^{\uparrow,\downarrow}=\frac{I_i^{\text{tot}} (1\pm P_i )}{2},
\end{equation}
where
\begin{math}
	i=\{x,y,z\}, \quad
	I_i^{\text{tot}}=(I_i^++I_i^-)
\end{math} and
$I_i^\pm$ is the measured intensity for the oppositely deflected electrons for a given channel, corrected by a relative efficiency calibration. The final spin polarisation is defined as follows

\begin{equation}
P_i=\frac{I_i^+-I_i^-}{S(I_i^++I_i^-)},
\end{equation}
where S is the Sherman function.

For the quantitative spin-polarisation magnitudes in Fig.~\ref{Fig4}, we fit the two peaks corresponding to to $v_1$ and $v_2$ to Lorentzian functions with a Shirley background. For each pair of EDCs, the peak widths and positions are held constant with only the intensity allowed to vary between the spin up and down channels. The spin-polarisation is then calculated from the relative areas of the Lorentzian peaks. The net polarisations from the two collections of bands, $v_1$ and $v_2$, are therefore extracted.  

\

\noindent{\textbf{Theoretical:}}

The electronic structure of HfSe$_2$ was calculated within DFT using the Perdew–Burke–Ernzerhof exchange-correlation functional~\cite{perdew1996}, modified by Becke–Johnson potential~\cite{tran2009} as implemented in the WIEN2K programme~\cite{wein2k}. Relativistic effects, including spin-orbit coupling, were fully included. The muffin–tin radius of each atom, $R_\text{MT}$, was chosen such that its product with the maximum modulus of reciprocal vectors $K_\text{max}$ become $R_\text{MT}K_\text{max} = 7.0$. The Brillouin zone was sampled by a $15\times15\times10$ $k$-mesh. To calculate the spin-projected band structures, we downfolded the DFT Hamiltonian into a 22-band tight-binding model using maximally localised Wannier functions~\cite{mostofi2008} with Hf-$5d$ and Se-$4p$ as the projection centers.

\end{document}